\documentclass[article,12pt,onecolumn,showpacs,preprintnumbers,amsmath,assume,plain]{revtex4}
\usepackage{graphicx}% Include figure files
\DeclareGraphicsExtensions{.jpg}
\usepackage{dcolumn}% Align table columns on decimal point
\usepackage{bm}% bold math

\def\bk{ {\bf k} }

\def\bd{ {\bf d} }
\def\be{ {\bf e_z} }

\def\re{ \,{\rm Re}\, }

\def\sr{ Sr$_2$RuO$_4$ }
\def\re{ \,{\rm Re}\, }

\def\be{ {\bf e_z} }
\def\sr{ Sr$_2$RuO$_4$ }

\def\re{ \,{\rm Re}\, }

\begin{document}

\title{A numerical calculation of the electronic specific heat for the compound \sr below its superconducting transition temperature}

\author{Pedro Contreras $^{1,2}$, Jos\'e Burgos $^{1}$, Ender Ochoa $^{1}$, Daniel Uzcategui $^{1}$}
\affiliation{$^{1}$ Departamento de F\'{\i}sica, Universidad de Los Andes, M\'erida 5101, Venezuela\\ $^{2}$ Centro de F\'{\i}sica Fundamental ULA}
\author{Rafael Almeida $^{3}$}
\affiliation{ $^{3}$ Departamento de Qu\'{\i}mica, Universidad de Los Andes, M\'erida 5101, Venezuela}
\date{\today}

\begin{abstract}
In this work, a numerical study of the superconducting specific heat of the unconventional multiband superconductor Strontium Ruthenate, \sr, is performed. Two band gaps models are employed, and the results rendered for each of them are compared. One of the models, previously proposed by one of the authors to explain the experimental temperature behavior of the ultrasound attenuation, considers two gaps with point nodes of different magnitude on different gap surface sheets, while the other one is an isotropic and line node model, reported in the literature for describing quantitatively experimental specific heat data. The \sr superconducting density of states DOS is computed by employing these two models and then, a detailed numerical study of the electronic specific heat, that includes the contribution from the different Fermi sheets, is carried out. It is found that the calculated point node model specific heat temperature behavior shows an excellent agreement with the existent \sr experimental data at zero field, particularly, it is obtained that the
observed specific heat jump at T$_c$ is precisely reproduced. Also, it is found that the sum of the contributions from the different bands fits quantitatively the measured specific heat data. The results in this work evidence that the \sr superconducting states are of unconventional nature, corresponding to those of a point node superconductor, and show the importance of taking into account the multiband nature of the material when calculating thermodynamic superconducting quantities.


\end{abstract}


\maketitle

\section{Introduction}\label{sec:intro}

The strontium ruthenate (\sr) is a multiband superconductor with Fermi surface composed of three sheets ($\alpha$, $\beta$ and $\gamma$ sheets).\sr has a body centered tetragonal structure with a layered square-lattice similar to that of many high temperature copper-oxide superconductors \protect\cite{mae94} and its normal state displays Fermi liquid behavior \protect\cite{ber3}. For pure samples, its critical temperature, T$_c$, is approximately 1.5 K, and is found that T$_c$ varies strongly with non magnetic impurity concentration. It has been proposed that \sr is an unconventional superconductor having some kind of nodes in the superconducting gap \protect\cite{mae94}. Thus, a number of theoretical works \protect\cite{zhi1,ric2,agt1} have predicted the existence of linear nodes on two of the three Fermi surface sheets ($\alpha$, $\beta$ sheets), while other ones  have proposed that the $\gamma$ sheet is nodeless \protect\cite{ric2,agt1}. The predictions in these works agree with the results obtained from measurements of specific heat $C(T)$ \protect\cite{deg1,deg2}, electronic heat transport $\kappa(T)$ \protect\cite{mak1}, and depth penetration  $\lambda(T)$ \protect\cite{bon1}. In electronic thermal conductivity and specific heat experiments, the three sheets have similar contributions to the $\kappa(T)$ and $C(T)$ results, i.e. they have an integral effect. Because of this, from these experiments is very difficult to discern if the order parameter in each of the Fermi sheets has similar nodal structure. In contrast, from \sr sound attenuation experiments \protect\cite{lup1} is possible to distinguish the nodal structure of the $\gamma$ sheet from those of the $\alpha$ and the $\beta$ sheets. Moreover, experiments on \sr ultrasound nodal activity $\alpha(T)$, measured below T$_c$, have yielded the anisotropy inherent to the $\bk$-dependence of electron-phonon interaction \protect\cite{wak4}. The results have showed that the $\gamma$ sheet dominates the ultrasound attenuation $\alpha(T)$ for the $L[100]$, $L[110]$ and $T[110]$ sound modes, and  that below T$_c$, these three modes exhibit comparable temperature power law behavior. These results lead to think that the $\gamma$ sheet and the experimentally measured ultrasound nodal activity should have nodal structure alike, which is similar to that displayed by the other two sheets, conclusion that contradicts the proposition of a nodeless $\gamma$ sheet. Additionally, according to \protect\cite{luk1}, the symmetry of the gap structure is believed to be a time reversal broken state, with the symmetry transforming as the two dimensional irreducible representation $E_{2u}$ of the tetragonal point group $D_{4h}$.

An extensive experimental investigation of the electronic specific heat $C(T)$ for the unconventional superconductor \sr has been performed in a series of experiments by Maeno and Collaborators \protect\cite{deg1,deg2,nis1}. Through these experiments, they look to elucidate the gap structure of this material by means of electronic specific heat measurements. Among these experiments, the Nishizaki and collaborators specific heat measurements \protect\cite{nis1} on clean samples of \sr and under zero magnetic field, showed a remarkable near-linear behavior of $C_{s}(T)/T$ at low temperatures. This result provides evidence, supporting the idea that \sr  is an unconventional superconductor with some kind of line of nodes in its order parameter. They pointed out that the measurements results were no consistent with a single band isotropic model with triplet order parameter, $d_z(k)$ $=$ $\Delta(T) (k_x+ik_y)$. As a consequence of this, any fittings performed with a single line node order parameter, or with multiband gaps having same order parameter on each band, will not agree with the experimental results. On the other hand, several theoretical works
\protect\cite{zhi1,nom1,ane1} have proposed models for calculating zero field specific heat, and their results have been able to successfully fit the experimental data. However, due to its relevance to our work, we will only refer to the calculation by Zhitomirsky and Rice
\protect\cite{zhi1}, which uses which uses a \sr nodeless $\gamma$ sheet superconductivity tight binding microscopic model. These authors performed a three parameter fitting to  specific heat experimental results \protect\cite{nis1}, reproducing well the experimental curve, but only rendering an approximate adjustment to the observed specific heat jump at T$_c$. Their calculations, that employed a lines of nodes model, yield a jump larger than that expected if a multiband model would have been used.

Recently, one of us has proposed a model based on symmetry considerations \protect\cite{wak4}, which is able to explain the experimental temperature behavior of the ultrasound attenuation for the $L[100]$, $L[110]$, and $T[110]$ sound modes \protect\cite{lup1}. According to this model, the $\gamma$ sheet should have well-defined point nodes, and the $\alpha$ and $\beta$ sheets could have also point nodes, but an order of magnitude smaller than those of the $\gamma$ band, and which resembles line of nodes with very small gap. In this work, this anisotropic model will be applied to calculate the specific heat will, aiming to improve the calculated value of the specific heat at T$_c$ . Summarizing, at this point there is a considerable consensus regarding the \sr unconventional superconducting behavior \protect\cite{mae94,ber3,ric2}, about the symmetry of the superconducting gap \protect\cite{luk1}, and, also about the multiband nature of the superconducting state; nevertheless, certainly, yet there is no agreement regarding the nodal structure of the superconducting gap on the different sheets of the Fermi surface. Within this context, taking into account the suitability shown by the gap model introduced in \protect\cite{wak4} to interpreting the electronic heat transport results below the transition temperature \protect\cite{con1}, in this article we will apply this model to the study of the electronic specific heat of \sr.

\section{The superconducting gap structure model}\label{sec:model}

As was mentioned before, here the gap model proposed in reference \protect\cite{wak4} will be extended to study the electronic specific heat. This model assumes a superconducting order parameter based on symmetry considerations, where the gap $\Delta_k^i (T)$ is given by:
\begin{equation}
\label{delta}
\Delta^{i}_k = (\bd^i (\bk) \cdot \bd^{i,\ast} (\bk) ) \; \Delta^i(T),
\end{equation}		
here $\Delta^i(T)$ is taken as $\Delta_0^i~\sqrt{1 - (T/T_c)^2}$, where $\Delta_0^i$ is an adjustable parameter from experimental data \protect\cite{mae2}. Before continuing it is important to point out that the particular choice for  $\Delta^i(T)$ does not seem to affect the final results. Thus, in this expression we have employed  $(T/T_c )^2$ instead of $(T/T_c)^3$  with no effect on the fitting of the experimental specific heat. The functions $\textbf{d}^i (k)$ are the vector order parameters for the $i$-Fermi sheets, transforming according to the two dimensional irreducible representation $E_{2u}$ of the tetragonal point group $D_{4h}$. The form of this function is \protect\cite{ric2},
\begin{equation}
            \bd^i (\bk) = \be [ d_x^i (\bk) + i \: d_y^i (\bk) ].
            \label{gap}
\end{equation}

Here $d_x^i$ and $d_y^i$ are real functions given by:
\begin{equation}
        \label{real_gap}
            d_x^i (\bk) = \delta^i\sin (k_x a) + \sin(\frac{k_x a}{2})
                \cos(\frac{k_y a}{2})\cos(\frac{k_z c}{2}),
\end{equation}
and
\begin{equation}
        \label{imag_gap}
            d_y^i (\bk) = \delta^i\sin (k_y a) + \cos(\frac{k_x a}{2})
                \sin(\frac{k_y a}{2})\cos(\frac{k_z c}{2}).
\end{equation}
In reference \protect\cite{wak4}, the factors $\delta^i$ were obtained by fitting the experimental data obtained from ultrasound attenuation measurements \protect\cite{lup1}.

\begin{figure}
\includegraphics[width = 2.6 in, height= 2.5 in]{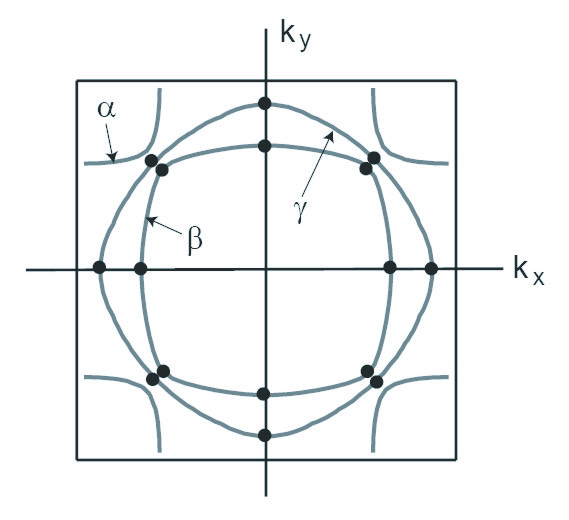}
\caption{\label{FS}The black dots show the
positions of the point nodes in the superconducting gap on the
$\beta$ and $\gamma$ Fermi surface sheets in \sr, as determined by
Eqs.~\ref{real_gap} and \ref{imag_gap}. Each solid circle
represents two nodes, at positions $\pm k_z$.}
\end{figure}

\begin{figure}
\includegraphics[width = 3.3 in, height= 3.8 in]{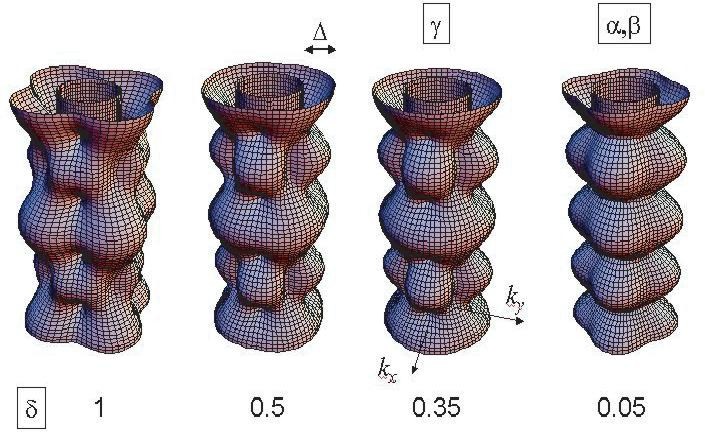}
\caption{\label{FS2}Three dimensional position of the point nodes in the superconducting gap on the $\alpha$ and $\beta$ and $\gamma$ Fermi surface sheets in the compound \sr as determined by Eqs. ~\ref{delta}, ~\ref{gap}, ~\ref{real_gap}, and ~\ref{imag_gap} for different values of the material parameter $\delta^i$. Each solid cone represents two nodes, at positions $\pm k_z$ \protect\cite{con2}.}
\end{figure}

For the $\gamma$ band nodal structure, this model predicts eight symmetry-related k-nodes lying on the symmetry equivalent ${100}$ planes (see Fig.~\ref{FS}), and also eight symmetry-related nodes in the ${110}$ planes. Similarly, for the $\alpha$ and $\beta$ sheets, the nodal structure of the order parameter yields eight symmetry-related $\textbf{k}$-nodes, lying on the symmetry equivalent ${100}$ planes (see also Fig.~\ref{FS}), and also eight symmetry-related nodes in ${110}$ planes. All these point nodes are "accidental" in the sense that they are not required by symmetry; instead, they only exist for certain range of the values of $\delta^{\gamma}$ and $\delta^{\beta/\alpha}$ material parameters \protect\cite{wak4}. The 3-dimensional results for the superconducting band structure of the $\alpha$, $\beta$ and $\gamma$ Fermi surface sheets, resulting by applying Eqs.~\ref{delta}, ~\ref{gap}, ~\ref{real_gap}, and ~\ref{imag_gap} for different values of the parameter $\delta^i$, are displayed in Fig.~\ref{FS2} \protect\cite{con2}.

The gap structure of \protect\cite{zhi1} can be described as follows: for the gamma sheet it has the form $d_z^\gamma(\bk) \propto c^\gamma (\sin k_x + i
\sin k_y)$ and for the $\beta$ and $\alpha$ sheets $d_z^{\beta/\alpha}(\bk) \propto c^{\beta/\alpha} (\sin k_x/2 \cos k_y/2 + i \cos k_x/2 \sin k_y/2) \cos (k_z c/2)$. The coefficients c$^\gamma$ and c$^{\beta/\alpha}$ are temperature dependent
quantities which fix the values for the maximum gaps $\Delta^\gamma_0$ $=$ c$^\gamma(T=0)$ and $\Delta^{\beta/\alpha}_0$ $=$ c$^{\beta/\alpha}(T=0)$. The order parameter for the $\gamma$ sheet in this model is nodeless, but the order parameter for the $\beta/\alpha$ sheets has horizontal line nodes located at k$_z c = \pi/ 2$.

\section{Superconducting density of states for a two gap model with point nodes of different magnitude}\label{density}

In this section, the results for the superconducting density of states are presented. Two models are employed; the first one considers two gaps with point nodes of different magnitude on different sheets of the Fermi surface \protect\cite{wak4}, while the second one, the Zhitomirsky and Rice model \protect\cite{zhi1}, assumes horizontal line nodes. In unconventional superconductors, the order parameter goes to zero at some parts of the Fermi surface. Due to this fact, the density of states at very low energy arises from the vicinity where the nodes of the order parameters are located. Well known examples of this are the high temperature superconductors. In general, line nodes and point nodes give a density of states that varies, at the low energy limit, as $\epsilon$ and $\epsilon^2$ respectively \protect\cite{min1}. Besides the nodes in the order parameter, scattering from non-magnetic impurities also influences the calculation of the low energy density of states \protect\cite{min1, hir1}. This scattering mechanism leads to the lowering of T$_c$ ; and therefore, to the suppression of the superconducting state. In general, for temperatures much smaller than T$_c$, the effect of very low concentrations nonmagnetic impurities can be neglected. It is found that only for very low temperatures, the effect of impurities becomes important for the so called unitary limit. However, for clean samples this effect can be neglected \protect\cite{min1,hir1}.

For the calculations of the density of states carried out in this work, we consider the tight-binding approximation to be valid. Hence, in the performed calculations we neglect any self-consistency, only the tight-binding structure of the normal state energy is considered and the order parameters are taken into account. Following the general approach, the unconventional superconductor Fermi surface-averaged density of states (DOS) can be calculated using the equation,
\begin{equation}
N^i (\epsilon) = N^i_0 \; \re [ \; g^i(\epsilon) \; ].
\label{ds1}
\end{equation}

Here, the $i$ label denotes the conduction bands $\alpha/\beta$ and $\gamma$. The quantity $N^i_0$ is the normal metal DOS at the Fermi level. For the case of a multiband superconductor, the following  $g^i(\epsilon)$ function is employed \protect\cite{con2,wak2},

\begin{equation}
g^i(\epsilon) = \left\langle \frac{\epsilon}{\sqrt{\epsilon^2-|\Delta^{i}_{k}|^2}}
    \right\rangle_{i_{FS}},
\label{ds2}
\end{equation}
where $\left\langle \cdots \right\rangle_{i_{FS}}$ denotes the average over the $i^{th}$ Fermi sheet. The numerical parameters involved in the tight-binding normal state energy are also used for calculating the superconducting DOS. Their values are determined from the band structure expression of the Fermi velocity, and correspond to  those in the Haas-Van Alfen and ARPES experiments \protect\cite{ber3,maz1}: $(E_0-E_F,t,t^\prime)$ $=$ $(-0.4,-0.4,-0.12)$.
The calculations performed here are for zero temperature systems, where  $\Delta_0^i$ is the zero amplitude gap parameter for the $i^{th}$ Fermi sheet. The normalized total density of states can be written in a dimensionless form as

\begin{equation}
\frac{N(\epsilon)}{N_0} = p^\gamma N^\gamma\; \Big(\frac{\epsilon}{\Delta^\gamma_0}\Big) + p^{\alpha \beta} N^{\alpha/\beta} \; \Big(\frac{\epsilon}{\Delta^{\alpha/\beta}_0}\Big).
\label{ds3}
\end{equation}

\begin{figure}
\begin{center}
\includegraphics[width = 3.0 in, height= 3.2 in]{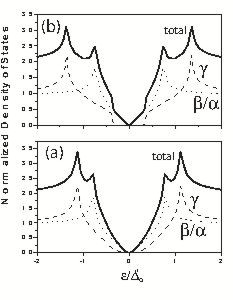}
\end{center}
\caption{\label{3} Normalized total and partial superconducting density of states
$N(\epsilon)$ for the multiband models with accidental point nodes
of \protect\cite{con1} panel (a), and the horizontal lines model
of \protect\cite{zhi1} panel (b).}
\end{figure}

Here $p^{\alpha \beta}$ and $p^\gamma$ are the fractions of the normal-state density of states in the normal metal associated with the $\alpha$, $\beta$ and $\gamma$ bands, respectively. These two quantities are related by the sum rule: $ p^\gamma + 2p^{\alpha \beta} = 1$. For our calculations, the experimentally determined values \protect\cite{ber3}, $p^{\alpha + \beta} = 0.42$ and $p^{\gamma} = 0.58$, are used. The results obtained for the Fermi averaged and normalized total DOS, together with those for each Fermi surface sheet, all at T $=$ 0 K, are displayed in Fig.~\ref{3}. The results in Fig. 3(a) correspond to the point node model \protect\cite{wak4}, while those in 3(b) are obtained by employing the Zhitomirsky and Rice lines of nodes model \protect\cite{zhi1}. For the point node model, it can be seen that the lines of very small point gaps on the $\alpha$ and $\beta$ sheets dominate the low energy behavior of the density of states. For these cases, the density of states increases faster than the density of states corresponding to the point gaps of the $\gamma$ sheet. This result agrees with that previously reported in \protect\cite{wak4}. The parameters $\Delta_0^{\beta/\alpha}$ and $\Delta_0^{\gamma}$ are determined in \protect\cite{mak1,con1} by fitting the thermal conductivity data, and their values will be also used for fitting $C_s(T)$. From the lines of nodes model, figure 3(b), one notices that the horizontal gaps on the $\alpha$ and $\beta$ sheets dominate the low energy behavior of the density of states, while the density of states of the $\gamma$ band opens a gap below certain critical value due to its nodeless nature of this model. For both cases, it is found that as more lines of nodes or point of nodes are added to the gap functions, the density of states inside the gap increases faster \protect\cite{wak4}. Also it is obtained that the density of states increases faster than linearly inside the superconducting region. In addition, both models present DOS van-Hove singularities, which are due to the tight-binding structure of \sr. These singularities are responsible for the two coherence peaks observed in the total density of states.

\section{Electronic superconducting specific heat in a multiband \sr}\label{sh}

\begin{figure}[t]
\begin{center}
\includegraphics[width = 3.0 in, height= 3.2 in]{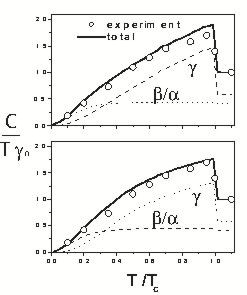}
\end{center}
\caption{\label{spheated}
Normalized electronic specific heat $C_s(T)/T$ for the multiband models: Lower panel displays the results yielded by the point node model \protect\cite{wak4,con1}, and the upper panel exhibits those obtained from the horizontal lines node model \protect\cite{zhi1}. The points correspond to the experimental data of \protect\cite{nis1}}.
\end{figure}

For an unconventional single band superconductor, the electronic specific heat low temperature behavior is expected to vary as $T^2$, for the horizontal lines of nodes (zeros) and as $T^3$ for the point of nodes models \protect\cite{min1}. For multiband superconductors, at $T$ smaller than T$_c$, the values of the specific heat show a strong dependence on the order parameters $\Delta^i$. However, due to the absence of a self-consistent evaluation of $\Delta^i(T)$ in the tight-binding method employed here, and since at temperatures close to T$_c$, the numerical value of the derivatives $ \frac{d \Delta^i(T)}{ dT}$ are  relevant, at first sight, a good agreement with the observed electronic specific heat behaviors may not be anticipated. Nevertheless, the order parameters have a strong dependence on the anisotropic effects \protect\cite{maz1,hir1,con2}. Thus, in this section we will explore if, through calculations that incorporate anisotropic effects, is possible to overcome the tight-binding method initial handicap, to provide a good description of the specific heat temperature behavior.  Before continuing it is important to point out that, as was done in the thermal conductivity case \protect\cite{con1}, for comparing the calculation results with the specific heat experimental data, the sum of the contributions from all the band sheets has to be considered.
The expression used in this work to calculate the electronic specific heat for an unconventional multiband superconductor is an extension of that developed in \protect\cite{hir1,dge1}, obtained through a single band unconventional superconductor formalism. Here, a more general expression for $C_{s}(T)$, that takes into account the anisotropic and multiband effects of an unconventional superconductor is used:

\begin{equation}
        \label{ce1}
            C(T)= \frac{2}{T} \sum_i \int d \epsilon \Big(-\frac{\partial f}{
            \partial \epsilon}\Big) \; F^i(\epsilon),
\end{equation}
the Fermi surface averaged function $F^i(\epsilon)$ appearing in this expression is given by:
\begin{equation}
 F^i(\epsilon) = p^i \; \left\langle N^i(\epsilon,\bk) \Big[ \epsilon^2 - T \;
 \frac{\mathrm{d} \; |\Delta^{i}_{k}|^2}{\mathrm{d} \; \mathrm{T}} \Big]\right\rangle_{i_{FS}};
\label{ca2}
\end{equation}
where $p^i$ denotes the partial density of states in the band $i$, $N^i(\epsilon,\bk)$) is the expression for the momentum dependence of the DOS and the quantity $N^i_0$ is the DOS at the Fermi level. The only input parameters required for the specific heat temperature behavior fitting are the zero temperature energy gaps $\Delta_0^\gamma$ and $\Delta^{\alpha/\beta}$ . Here we use the values calculated in \protect\cite{mak1,con1} for fitting the thermal conductivity data. For the models employed here, the accidental point node and the horizontal line node models, fig.~\ref{spheated} shows the theoretical results for the temperature dependence of the normalized electronic specific heat, calculated from Eq.\ref{ce1} together with the experimental results reported in \protect\cite{nis1}. For the point node model \protect\cite{wak4,con2,con1}, the lower panel displays of fig.~\ref{spheated}  shows an excellent fitting of the experimental data.

One point which shows excellent agreement is the size of the specific heat jump at
T$_c$, $\Delta C/ C_n$ $\simeq$ 0.73 for our point model \protect\cite{wak4}, while that for the line nodes model model of \protect\cite{zhi1} the value is $\Delta C/ C_n$ $\simeq$ 0.82. The experimental work of Nishizaki \protect\cite{nis1} gives $\Delta C/ C_n$ $=$ 0.70. This is a remarkable result since, as mentioned before, only anisotropic arguments are considered, and also because only two parameters, obtained from the literature, corresponding to fitting a different thermal conductivity experiment, are used. The contributions of each of the three band sheets are displayed in the figure, from there is observed that at low temperatures, close to T$=$0, the anisotropy of the gap in the $\alpha$ and $\beta$  bands dominates the behavior, while at higher temperatures, the electronic specific heat behavior is dominated by the $\gamma$ band. The results in the lower panel of fig.~\ref{spheated} show that the horizontal line node model also provide a good fitting; however, a worse adjustment to the experimental data is obtained as T approaches T$_c$, and, as reported in the literature \protect\cite{zhi1,ber3,mae2} , from this model, the theoretical calculations only yield a modest adjustment to the observed specific heat jump at T$_c$ i.e. this jump seems to be better reproduced by our point nodes model. It is interesting to point out that
\protect\cite{deg1,nis1}reported that the $\gamma$-band is the one responsible for the largest contribution to the total density of states; thus, as a consequence of this, Fig.~\ref{spheated} also shows that, for both models, the $\gamma$-band dominates the contribution to the specific heat.

\section{Conclusions}\label{conclu}

In this work, two different models for the gap structures characterizing the $\alpha$, $\beta$ and $\gamma$ Fermi surface sheets are employed to calculate the \sr superconducting electronic specific heat $C_{s}(T)$. One of them considers two gaps with point nodes of different magnitude on different sheets, while the other one assumes horizontal line nodes. Through the first model, it is found that the calculated temperature behavior of $C_{s}(T)$ shows  an excellent agreement with the existent \sr experimental data at zero field ~\protect\cite{deg1,deg2,nis1}, particularly, the observed specific heat jump at T$_c$ is better reproduced by the point nodes model. The results obtained here seem to confirm that the \sr superconducting state corresponds to that of a point node unconventional superconductor.

\section*{Acknowledgments}

We thank Dr. Y. Maeno for providing the experimental data in Fig.~\ref{spheated}. We also acknowledge discussions with Prof. Michael Walker from the University of Toronto, and Profesores Luis Rinc\'on and Andr\'es Eloy Mora from Universidad de Los Andes. This research was supported by the Grant CDCHTA number C-1851-13-05-B.

%===================================================================

\end{document}